\documentclass{article}
\usepackage{tikz}

\usepackage[preprint]{neurips_2024}

\usepackage[utf8]{inputenc} 
\usepackage[T1]{fontenc}    
\usepackage{hyperref}       
\usepackage{url}            
\usepackage{booktabs}       
\usepackage{amsfonts}       
\usepackage{nicefrac}       
\usepackage{microtype}      
\usepackage{xcolor}         

\usepackage[strings]{underscore}
\usepackage[utf8]{inputenc} 
\usepackage[T1]{fontenc}    
\usepackage{hyperref}       
\usepackage{url}            
\usepackage{booktabs}       
\usepackage{amsfonts}       
\usepackage{nicefrac}       
\usepackage{microtype}      
\usepackage{xcolor}         
\usepackage{graphicx}
\usepackage{bbding}
\usepackage{multirow}
\usepackage{graphicx}
\usepackage{colortbl}  %
\usepackage{xcolor}    %
\usepackage{subcaption}
\usepackage{tabularx}
\usepackage{lipsum}
\definecolor{lb}{HTML}{D8F3FD} 
\usepackage[listings]{tcolorbox}
\tcbuselibrary{listings,theorems}
\lstdefinestyle{go}{
    language=Go,
    frame=lines,
    basicstyle=\ttfamily\small,
    keywordstyle=\color{blue},
    commentstyle=\color{teal},
    stringstyle=\color{red},
    numbers=left,
    numberstyle=\tiny\color{gray},
    stepnumber=1,
    numbersep=5pt,
    showstringspaces=false,
    breaklines=true,
    breakatwhitespace=true,
    tabsize=4,
    captionpos=b
}
\definecolor{purplex}{HTML}{9564bf}
\definecolor{blue1}{HTML}{508AB2}
\newtcbtheorem[]{exmp}{Example}%
{colback=gray!5, colframe=gray!80, fonttitle=\bfseries, 
 width=0.47\textwidth, 
 left=0.05in, right=0.05in, bottom=0.05in, top=0.05in}{exmp}

\title{Tool-integrated Reinforcement Learning for Repo Deep Search}

\author{Zexiong Ma\thanks{\, Work done during the internship at ByteDance.} \hspace{0.4mm}$^{\heartsuit}$,\, Chao Peng\thanks{\, Corresponding authors.} \hspace{0.4mm}$^{\clubsuit}$, Qunhong Zeng$^{*\diamondsuit}$,\, Pengfei Gao$^{\clubsuit}$,\\\textbf{Yanzhen Zou}$^{\heartsuit}$,\, \textbf{Bing Xie}$^{\dagger\heartsuit}$
\vspace{1mm}\\
  $^{\heartsuit}$Peking University,\,\,
  $^{\clubsuit}$ByteDance,\, 
  $^{\diamondsuit}$Beijing Institute of Technology\vspace{1mm}\\
  $^{\heartsuit}$\texttt{\{mazexiong@stu., zouyz@, xiebing@\}pku.edu.cn},\\
  $^{\clubsuit}$\texttt{\{pengchao.x, gaopengfei.se\}@bytedance.com}, \\
  $^{\diamondsuit}$\texttt{qunhongzeng@bit.edu.cn}
}

\begin{document}

\maketitle

\begin{abstract}
Issue localization, the process of identifying code locations that need modification to resolve software issues, is a critical yet challenging task in software development. The semantic gap between natural language issue descriptions and faulty code requires complex multi-hop reasoning through code dependencies. Existing LLM-based agents attempt to address this by integrating repository retrieval tools. However, this transforms issue localization into a demanding task we call \textit{Repo Deep Search}, which requires the LLM to effectively utilize various repository retrieval tools throughout a multi-step reasoning and navigation process. To tackle this challenge, we present \textit{ToolTrain}, a two-stage tool-integrated training framework combining rejection-sampled supervised fine-tuning and tool-integrated reinforcement learning to enhance LLMs' ability to use retrieval tools for issue localization. Experimental results show that ToolTrain-trained models achieve state-of-the-art performance, with our 32B model even surpassing Claude-3.7 on function-level localization. The results also show that improved localization performance translates to better end-to-end issue resolution performance. This further demonstrates that training for issue localization is a viable and effective strategy for improving automated software development.
\end{abstract}

\begin{center}
    \vspace{-5mm}
    \fcolorbox{white}{white}{\parbox{.99\linewidth}{
    \centerline{\textit{We shape our tools, and thereafter our tools shape us.}}
    \rightline{\textit{---Marshall McLuhan}}
}}
\end{center}

\section{Introduction}
Issue localization is the process of identifying the exact locations in source code that need to be modified to resolve a software issue \citep{yang2024swe}. Given an issue description (e.g., a bug report in natural language), developers must navigate through the entire repository to locate the specific files, functions, or code segments that require changes~\citep{wen2016locus}.
Issue localization is a time-consuming and labor-intensive task, particularly for complex, large-scale repositories~\citep{zhu2022enhancing, youm2017improved}. This has motivated growing research into automated approaches, and with LLMs' remarkable success in software engineering tasks such as code generation \citep{jiang2024survey, austin2021program, li2022competition} and test synthesis \citep{wang2024hits, fakhoury2024llm}, LLM-based issue localization has emerged as a natural and promising direction.

LLM-based agents~\citep{chen2025locagent,yang2024swe} allow LLMs to use various tools (e.g., searching specific function by keyword, listing functions that call a certain function) to dynamically explore repositories. However, this approach places higher demands on LLMs: beyond multi-step reasoning, they require sophisticated tool-calling abilities to explore code effectively rather than walking aimlessly through the repository. We refer to this complex, sequential navigation and discovery task as \textbf{\textit{Repo Deep Search}}. Current LLMs often struggle with these high-demand requirements, making incorrect tool calls or failing to maintain coherent reasoning chains throughout the exploration process. Inspired by recent advancements in information retrieval where Reinforcement Learning (RL) agents are trained for multi-step \textit{Deep Search}~\citep{deepresearch, jin2025search}, we investigate whether this paradigm can be extended to \textit{Repo Deep Search}. Specifically, we explore using RL to enhance an LLM's autonomous navigation within a code repository, aiming to significantly improve its performance on what we define as \textit{Repo Deep Search} for issue localization.

In this paper, we propose \textbf{\textit{ToolTrain}}, a tool-integrated training framework to enable LLMs to perform more effective multi-hop reasoning with tool call during issue localization. Firstly, we implement \textbf{\textit{RepoSearcher}}, a lightweight and LLM-friendly issue localization agent. \textit{RepoSearcher} includes simple and easy-to-use retrieval tools that allow LLMs to retrieve function or class definitions by name. To help the LLMs better leverage these tools for multi-hop reasoning, we construct a set of labeled issue localization training data from open-source repositories and train the LLMs in two stages: (1) \textbf{Rejection-sampled supervised fine-tuning (SFT)}, where the LLM generates tool-use trajectories, and only those leading to correct answers are used for fine-tuning. This stage serves as a warm-up to help the LLM learn the task format and how to call the retrieval tool. (2) \textbf{Tool-integrated reinforcement learning (RL)}, where the LLM learns through trial and error by exploring different tool-use strategies. We employ rule-based reward signals that evaluate whether the LLM's exploration successfully locates the correct code elements. This direct feedback teaches the model which tool-calling patterns lead to successful localization and which lead to dead ends. Through this iterative process, the LLM learns to make more strategic tool calls at each step, avoiding redundant explorations and focusing on promising code paths. This two-stage training enables the LLMs to master multi-hop reasoning with retrieval tools, yielding more precise issue localization.

We applied \textit{ToolTrain} to train LLMs and compared \textit{RepoSearcher} with ToolTrain-model against multiple issue localization frameworks~\citep{jiang2025cosilsoftwareissuelocalization,agentless,chen2025locagent,yu2025orcalocallmagentframework}. The results demonstrate that our method achieves state-of-the-art (SOTA) performance on the issue localization task among same-size LLMs, and on some function-level localization tasks, it even surpasses leading LLMs like Cluade-3.7.
We also experimentally validated that more accurate issue localization results lead to better issue resolution performance. Furthermore, through a detailed analysis of the \textit{ToolTrain} process, we found that reinforcement learning (RL) effectively enhances the LLM's tool-calling and reasoning abilities. This enables the model to explore the repository with greater efficiency and precision, allowing it to accurately localize the faulty position.

In summary, this paper makes the following main contributions:
\begin{itemize}
    \item We design \textit{RepoSearcher}, a lightweight issue localization agent, to streamline the training of issue localization agents.
    \item We propose \textit{ToolTrain}, a two-stage tool-integrated training approach, to enhance the reasoning capabilities of LLMs during the tool-calling process.
    \item We use \textit{ToolTrain} to train open-source LLMs and validate the effectiveness of \textit{RepoSearcher} with ToolTrain-model.
\end{itemize}

\section{Background and Motivation}\label{sec:background}
In this section, we first introduce issue localization task and issue localization agent. Then we present two LLM post-training techniques: rejection-sampled supervised fine-tuning and rule-based reinforcement learning. 

\subsection{Issue Localization}

Issue localization~\citep{chen2025locagent} refers to the task of identifying faulty file or code snippets in a repository based on natural language issue descriptions (e.g., bug reports). Issue localization is a critical step in the automated issue resolution pipeline, as accurate localization can significantly reduce the time and effort developers spend understanding and fixing problems. However, issue localization poses several challenges. On one hand, there is often a significant semantic gap between the issue description and the actual faulty code. The issue typically describes user-facing symptoms, while the root cause may lie deep in the underlying implementation~\citep{li2021practical}. On the other hand, modern software systems are large and structurally complex, making exhaustive code search both inefficient and impractical.

To address these challenges, recent studies have explored LLM-based issue localization agents~\citep{chen2025locagent,jiang2025cosilsoftwareissuelocalization, yu2025orcalocallmagentframework}, which allow LLMs to interact with code retrieval tools and perform multi-turn information gathering and reasoning, enabling more precise issue localization. The core idea behind these approaches is to empower LLMs with the ability to actively issue retrieval commands, dynamically query relevant parts of the repository based on context, and progressively narrow down the search space. However, existing LLMs are generally not trained specifically for tool-augmented reasoning, which limits their efficiency in using tools for issue localization. This problem is particularly pronounced when the issue descriptions are incomplete or ambiguous, the model often struggles to accurately understand the problem and make successful tool calls. Therefore, a key research challenge is to \textbf{enhance the tool call accuracy of LLMs and their ability to reason over retrieved information during issue localization}.

\subsection{LLM Post-Training}
LLM post-training refers to the further training LLMs with (question, answer) training data after the pretraining phase. Post-Training aims to enhance the model's performance in specific tasks or usage scenarios. In this section, we will introduce two mainstream LLM post-training techniques: rejection sampled supervised fine-tuning and rule-based reinforcement learning.
\subsubsection{Rejection-Sampled Supervised Fine-tuning}

Rejection-sampled supervised fine-tuning (SFT)~\citep{pantraining} is a post-training approach that improves LLM's performance with a three-stage pipeline: sampling, filtering, and fine-tuning. First, the LLMs generate multiple candidate responses for a given question. Next, a filtering mechanism evaluates these responses against a ground-truth answer, selecting only the high-quality responses. Finally, this curated set of question-answer pairs is used as the dataset for supervised fine-tuning.
In agent-based settings, an agent is used to generate multiple tool call trajectories. High-quality trajectories are selected for fine-tuning based on whether the final result matches the ground-truth answer. This approach could improve the model’s ability to call tools and complete tasks effectively.
However, a key limitation of SFT is that it only leverages successful (i.e., high-quality) trajectories during training, while ignoring the negative supervision signals embedded in failed trajectories. This one-sided training paradigm can lead the model to ``memorize'' superficial patterns from a small number of successful examples~\citep{chu2025sft}, rather than truly learning the reasoning logic behind effective tool use. As a result, the fine-tuned model may suffer from limited generalization, especially when faced with out-of-distribution tasks, complex reasoning paths, or novel tool combinations~\citep{schick2023toolformer}.

\subsubsection{Rule-based Reinforcement Learning}

Rule-based reinforcement learning (RL)~\citep{guo2025deepseek} involves a three-stage process: sampling, scoring, and training. Initially, the LLM samples a set of responses to a prompt. These responses are then evaluated against a ground-truth answer to produce a reward score. This score is subsequently used as the supervision signal to fine-tune the model via reinforcement learning. 
Rule-based reinforcement learning evaluates the quality of reasoning processes solely based on the correctness of the final answer, using reinforcement learning to encourage reasoning paths that lead to correct answers while discouraging those that produce incorrect ones. This approach has proven effective in enhancing LLMs' reasoning abilities in complex tasks~\citep{ma2025sorft}. Both OpenAI-o1~\citep{jaech2024openai} and DeepSeek-R1~\citep{guo2025deepseek} have employed RL to improve LLMs' reasoning capabilities.

In agent-oriented scenarios, rule-based reinforcement learning enables the agent to sample multiple reasoning trajectories (including tool invocations) and optimizes the entire trajectory through reinforcement learning based on scoring results. Several existing works have explored RL to improve LLM-based agents, such as OpenAI's Deep Research~\citep{deepresearch} and Operator~\citep{operator}, Anthropic's Claude Code~\citep{claudecode}, and open-source efforts like Search-R1~\citep{jin2025search}.
Compared to fine-tuning methods that only utilize high-quality trajectories, RL not only improves the model's ability to generate correct trajectories by reinforcing high-scoring paths, but also reduces the likelihood of low-quality outputs by penalizing incorrect trajectories through low scores.
By leveraging both positive and negative samples, RL makes more efficient use of the LLM's sampling outputs, significantly enhancing the model's generalization capabilities~\citep{xie2025logic, chu2025sft}.

\subsection{Motivation}

Issue localization is a critical task in software engineering that requires LLM-based agents to make complex reasoning, and calling various search tools to navigate repositories. However, current LLMs exhibit poor performance on this task due to the lack of tool-integrated training. Their ability to reason throughout the tool-calling process remains a significant bottleneck.
To address this gap, we explore how to effectively enhance the tool-calling capabilities of LLMs through post-training. A straightforward approach is rejection-sampled supervised fine-tuning (SFT), which can directly and simply improve an LLM's performance on issue localization. However, we contend that \textbf{excessive fine-tuning risks compromising the model's generalization capabilities}, potentially harming the broad reasoning skills that make LLMs powerful in the first place.
On the other hand, Rule-based Reinforcement Learning (RL) holds great potential for enhancing an agent's capabilities in issue localization. However, a major challenge lies in constructing suitable training data for an issue localization agent. Specifically, ensuring the reliability of ground-truth answers is a difficult and underexplored problem.
In this paper, we aim to tackle these challenges head-on. First, we will investigate methods for constructing high-quality training data specifically designed for the issue localization task. Second, we will explore how SFT and RL can be effectively combined. Our goal is to develop a hybrid training methodology that leverages the strengths of both approaches to significantly enhance the issue localization capabilities of LLM-based agents.

\section{Approach}\label{sec:approach}

In this section, we will introduce our approach, which consists of three components. (1) Issue Localization Agent: We designed \textit{RepoSearcher}, which contains a set of simple yet efficient tools for the issue localization task to reduce the difficulty for LLMs in acquiring repository context. (2) Training Data Construction: We constructed training data for issue localization based on (issue, pull request) data from open-source communities. (3) Tool-integrated Training: We propose \textit{ToolTrain}, leverages our constructed data to perform tool-integrated training for LLMs. This training process enables the models to efficiently use tools for reasoning, further improves their accuracy in issue localization.

\begin{figure*}[t]
  \centering
  \includegraphics[width=.95\textwidth]{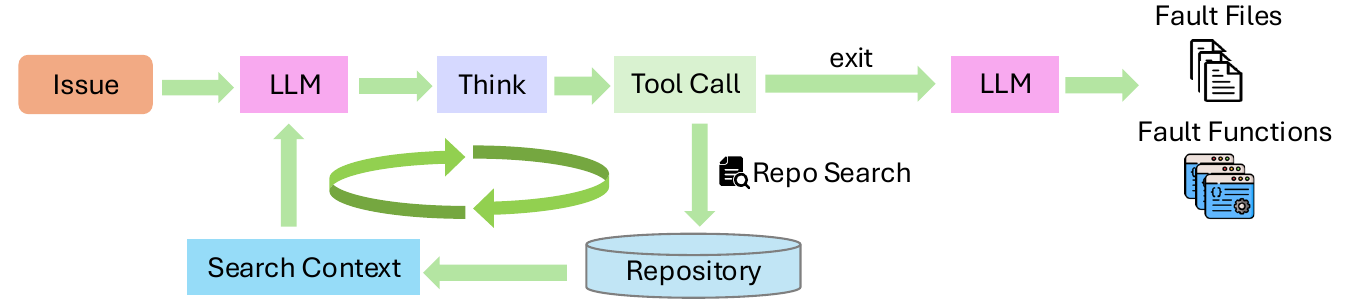}
  \caption{\textit{RepoSearcher} framework. Through multiple rounds of tool calls, \textit{RepoSearcher} navigates across different functions within to gather relevant information. Once LLMs determine that all necessary information is collected, it calls the \textit{Exit} tool and provides the final localization results.}
  \label{fig:approach}
\end{figure*}

\subsection{Issue Localization Agent}

An LLM-based agent capable of excelling at issue localization and amenable to tool-integrated training should be designed with the following characteristics: (1) \textbf{Well-Defined and Usable Tools}: The tools must be clearly defined, simple to use, and enable the LLM to accurately retrieve content from the target repository. (2) \textbf{Concise Trajectories}: The localization should not require an excessive number of tool calls. This is critical to prevent the generation of overly long trajectories, which cannot be fully utilized as complete samples for training the LLM.

As shown in Table~\ref{tab:tools}, we designed a set of simple and easy-to-use search tools for retrieving repository content. This suite includes six tools: \textit{GetRepoStructure}, \textit{GetImportOfFile}, \textit{SearchClass}, \textit{SearchFunction},  \textit{SearchClassMethod}, and \textit{Exit}.
The \textit{GetRepoStructure} and \textit{GetImportOfFile} tools are designed to quickly and accurately provide the repository's structural information and dependencies. This helps the LLM understand the overall repository architecture and perform coarse-grained localization. On the other hand, \textit{SearchClass}, \textit{SearchFunction}, and \textit{SearchClassMethod} provide the model with specific code snippets from the repository. This allows the LLM to inspect the content of specific functions and perform fine-grained localization. Once the LLM determines that it has gathered sufficient information, it invokes the \textit{Exit} tool to conclude the search process.

As shown in Figure~\ref{fig:approach}, we propose \textbf{\textit{RepoSearcher}}, a lightweight issue localization agent. Initially, \textit{RepoSearcher} generates a preliminary thought based on the input issue description and invokes a retrieval tool to fetch content from the repository. Based on the retrieved results, it proceeds with the next cycle of reasoning and tool invocation. Through multiple rounds of tool calls, \textit{RepoSearcher} navigates across different functions within the project to gather relevant information. Once the LLM determines that all necessary information has been collected, it calls the Exit tool and provides the final issue localization result based on the accumulated context.

\begin{table*}[t]
  \centering
  \caption{Repository search tools for \textit{RepoSearcher}.}
  \label{tab:tools}
  \scalebox{0.65}{
  \begin{tabular}{l|l|l}
      \hline
      \textbf{API name} & \textbf{Description} & \textbf{Output} \\
      \hline
      GetRepoStructure ( ) & Get the repository file structure. & The repository file structure. \\
      \hline
      GetImportOfFile (\texttt{file}) & Get the imports of given \texttt{file}. & The imports of \texttt{file}.   \\
      \hline
      SearchClass (\texttt{file}, \texttt{class}) & Search for the content of \texttt{class} in the \texttt{file}. & Code content of the searched class. \\
      \hline
      SearchFunction (\texttt{file}, \texttt{function}) & Search for the content of \texttt{function} in the \texttt{file}. & Code content of the searched function. \\
      \hline
      SearchClassMethod (\texttt{file}, \texttt{class}, \texttt{method}) &  Search for the content of \texttt{method} in the \texttt{class} of \texttt{file}. & Code content of the searched method. \\
      \hline
      Exit ( ) & Exit if LLM have found all the information needed. & - \\
      \hline
  \end{tabular}
  }
\end{table*}

\subsection{Training Data Construction}
In this paper, we construct training data for the issue localization task based on (issue, pull request) pairs from high-quality projects on GitHub. We select 600 high-quality repositories from GitHub according to the following criteria: (1) at least 1,000 issues; (2) at least 1,000 pull requests; (3) at least 100 stars; (4) inclusion of an appropriate license. To prevent data leakage, we excluded repositories and issues that appeared in SWE-Bench-Verified (our evaluation dataset, see Section~\ref{sec:benchmark} for details).

From these repositories, we construct issue localization training data by pairing resolved issues with their corresponding pull requests. To ensure data quality, we apply several filtering criteria: First, we discard issues with fewer than 100 characters to ensure sufficient descriptive content. Second, we filter out pull requests that only contain documentation updates, configuration changes, or other non-code modifications, focusing on modifications to the actual source code.
Through this selection process, we constructed around 28k high-quality localization examples. For each example, we form the question $q$ from the issue description and use the source code files and functions modified in the corresponding pull request as the ground-truth answer $A_q$ (excluding any non-source code elements). By leveraging authentic, high-quality, and successfully resolved issues as the foundation of our training data, we ensure both the relevance and reliability of our dataset, which in turn drives the effectiveness and stability of our training pipeline.

\begin{figure*}[t]
  \centering
  \includegraphics[width=.99\textwidth]{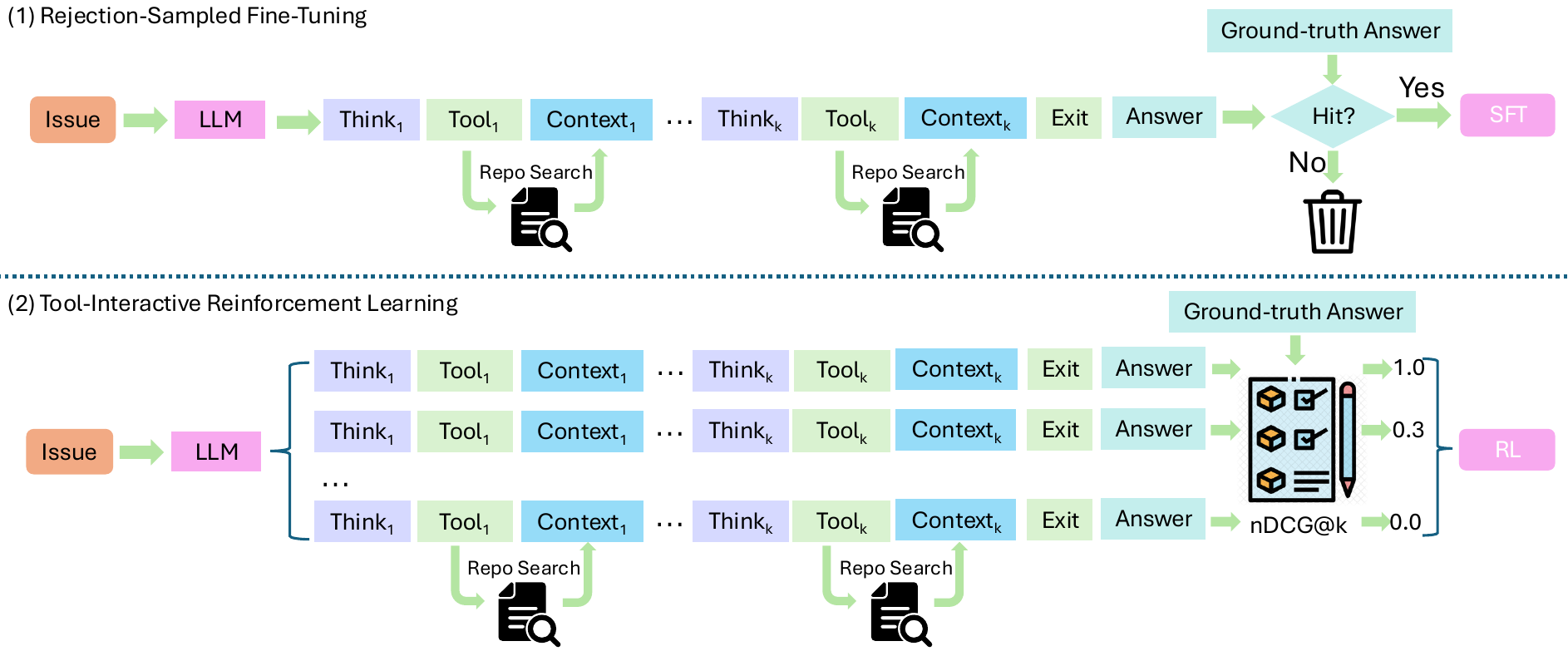}
  \caption{Tool-integrated training framework. It consists of two stages: (1) rejection-sampled supervised fine-tuning (SFT) and (2) rule-based reinforcement learning (RL).}
  \label{fig:tooltrain}
\end{figure*}

\subsection{Tool-integrated Training}
We propose \textbf{ToolTrain}, a two-stage tool-integrated training framework, to enhance the LLM's ability to reason and interact with external tools effectively. It contains two training stages: (1) Rejection-sampled supervised fine-tuning (SFT). We fine-tune the LLM exclusively on high-quality trajectories selected via rejection sampling. This process equips the model with a foundational understanding of task formats, tool invocation methods, and core reasoning strategies.
(2) Rule-based reinforcement learning (RL), we score the trajectories sampled by the LLM to increase the probability of it generating correct trajectories while simultaneously discouraging it from producing incorrect ones. This further improves the LLM's multi-hop reasoning capabilities during the tool-calling process.

\subsubsection{Rejection-Sampled Supervised Fine-Tuning}

We use \textit{RepoSearcher} to sample agent trajectories for issue localization task, and then select high-quality trajectories to fine-tune the LLM. More specifically, given the task description of issue, LLM invokes search tools to retrieve relevant context and generates a localization result. We then filter out low-quality trajectories that completely fail to hit the ground-truth answer, and use the remaining high-quality trajectories to fine-tune LLM. The main goal of this stage is to warm up LLM for tool use, enabling it to understand the task format of project retrieval, how to use the tools, and the basic reasoning strategies for issue localization.

\subsubsection{Tool-integrated Reinforcement Learning}

Although SFT can quickly teach LLM the strategy for issue localization, the fine-tuned LLM may simply ``memorize'' the correct reasoning paths, leading to poor generalization when faced with new issues. We employ tool-integrated reinforcement learning to robustly enhance the LLM's reasoning capabilities during tool interaction. More specifically, the LLM samples tool-use trajectories during the issue localization process, and each trajectory's predicted answer is scored based on its alignment with the ground-truth answer. These scores serve as a reward signal, guiding the model's update through reinforcement learning. By incentivizing high-quality trajectories and penalizing low-quality ones, this RL framework not only polishes the accuracy of tool interactions but also teaches the model to avoid erroneous or illogical tool calls. Consequently, it simultaneously enhances both the reliability and efficiency of the entire tool-use process.

We utilize nDCG@k (Normalized Discounted Cumulative Gain at rank k)~\citep{wang2013theoretical} to measure trajectory quality, as it holistically assesses both the recall of correct functions and their ranking position.
In the issue localization training process, the Agent is expected to output a ranked list of functions that are potentially related to a given issue. We are not only concerned with whether the Agent can cover all correct target functions, but also with whether it can rank these functions as high as possible in the list. This is crucial because, in practical scenarios, downstream systems (e.g., automated program repair agents~\citep{wang2024opendevin, yang2024swe, chen2024coder}) typically prioritize only the top-ranked candidate functions. If the correct functions are ranked too low, they may be ignored by repair agents.
Therefore, we use nDCG@k as the reward function to evaluate the ranking quality of the predicted list relative to the ground-truth set. nDCG@k is calculated by Equation~\ref{eq:ndcg}, it assigns different gains based on the positions of correctly predicted functions in the output list, functions ranked higher contribute more to the final score. By incorporating nDCG@k as a reward signal in the reinforcement learning process, we effectively encourage the Agent not only to “find the correct functions” but also to “rank the correct functions higher,” which aligns more closely with the practical requirements of real-world issue localization scenarios.

\begin{equation}
\text{nDCG@}k(q) = \frac{\text{DCG@}k(q)}{\text{IDCG@}k(q)}
\label{eq:ndcg}
\end{equation}

\begin{equation}
\text{DCG@}k(q) = \sum_{i=1}^{k} \frac{\mathbb{I}(L_q[i] \in A_q)}{\log_2(i+1)}
\end{equation}

\begin{equation}
\text{IDCG@}k(q) = \sum_{i=1}^{\min(k, |A_q|)} \frac{1}{\log_2(i+1)}
\end{equation}

where \( L_q[i] \) is the item at rank \( i \) in the predicted list for query \( q \), \( A_q \) is the set of ground-truth relevant items for query \( q \), \( \mathbb{I}(\cdot) \) is the indicator function that returns 1 if the condition holds, and 0 otherwise.

\section{Experimental Setup}\label{sec:setup}
In this section, we will introduce our experimental setup, including: benchmark, evaluation metrics, baselines and implementation details.

\subsection{Benchmark}\label{sec:benchmark}
In this paper, we construct our evaluation dataset based on SWE-Bench-Verified\footnote{https://openai.com/index/introducing-swe-bench-verified/}.
\textbf{SWE-Bench-Verified} is a benchmark designed to evaluate the issue resolution capabilities of LLM-based Agents, built from real issues collected from GitHub. To ensure accurate and reliable evaluation of LLM-based Agents on issue resolution, OpenAI invited 93 professional software developers to analyze and manually verify 2,000 real issues from GitHub. Based on this effort, a curated set of 500 verifiably solvable issues was constructed to form the SWE-Bench-Verified dataset. It has been adopted by leading large language model providers (e.g., OpenAI~\citep{gpt4}, Qwen~\citep{hui2024qwen2}, and Anthropic~\citep{claude4}) to assess the performance of cutting-edge models in automatic software development tasks.
In this paper, we use the functions and files modified by the golden patches in SWE-Bench-Verified as the ground-truth answers for the issue localization task, to evaluate the performance of our \textit{RepoSearcher} with ToolTrain-model.

\subsection{Metrics}
To systematically evaluate the effectiveness of {RepoSearcher} with ToolTrain-model, we conducted a multi-dimensional evaluation using multiple metrics: Recall@k, MAP, MRR, nDCG@k and \%Resolved. See Appendix \ref{sec:detailsOfMetrics} for detailed description of metrics.

\subsection{Baselines}
To evaluate the effectiveness of \textit{RepoSearcher} with the ToolTrain-model on issue localization, we conduct comparisons against multiple issue localization frameworks across various models.
\subsubsection{Frameworks} We selected four state-of-the-art issue localization frameworks with different design philosophies as baselines: Agentless\citep{agentless}, CrcaLoca\citep{yu2025orcalocallmagentframework}, CoSIL\citep{jiang2025cosilsoftwareissuelocalization}, LocAgent\citep{chen2025locagent}. See Appendix \ref{sec:detailsOfBaseline} for detailed description of baseline frameworks.

\subsubsection{Models}
We applied \textit{ToolTrain} to the Qwen2.5-Coder-7B-Instruct (hereafter Qwen-7B) and Qwen2.5-Coder-32B-Instruct (hereafter Qwen-32B) models~\citep{hui2024qwen2} and compared them against several baselines. For frameworks without their own fine-tuned versions (Agentless, Orcaloca, and CoSIL), we used the off-the-shelf Qwen-7B and Qwen-32B models. For LocAgent, we used its specialized CL-7B and CL-32B models. Our evaluation also included a comparison between our ToolTrain-models and state-of-the-art proprietary models like GPT-4o and Claude-3.7-Sonnet\citep{claude37}.

\subsection{Implementation Details}
Among 28k issue localization examples, 10k examples were used to sample \textit{RepoSearcher} trajectories with Claude-3.7-Sonnet~\citep{claude37}. Low-quality trajectories were filtered based on ground-truth data, resulting in 5k high-quality trajectories for SFT. The remaining 18k examples were used for RL training.
We use verl \citep{sheng2025hybridflow} for training in both the SFT and RL stages. In the SFT stage, we set the $\mathrm{train\_batch\_size}$ to 128 and train for 3 epochs on the 5k training examples. For the RL stage, we adopt the SGLang\footnote{https://github.com/sgl-project/sglang} inference framework to accelerate the sampling process and set the temperature to 1.0 to encourage response diversity. We set $\mathrm{train\_batch\_size}$ to 128, with $\mathrm{max\_prompt\_length}$ set to 12k and $\mathrm{max\_response\_length}$ to 20k, and train for one epoch on 18k examples.

\section{Results and Analysis}\label{sec:results}
In this section, we will present the experimental results for the three research questions, and experimental analysis.

\begin{table*}[t!]
  \centering
  \caption{Issue localization results on SWE-Bench-Verified. \colorbox{blue!8}{Blue background} indicates the results of ToolTrain-model; \textbf{bold numbers} denote the best performance among same-size models, while \underline{underlined numbers} indicate the best performance across all models.}
  \resizebox{.99\linewidth}{!}{
  \begin{tabular}{c|c|cccccc|cccccc}
  \toprule
  \multirow{2}{*}{\textbf{Framework}} & \multirow{2}{*}{\textbf{Model}} & \multicolumn{6}{c|}{\textbf{File-level}}  & \multicolumn{6}{c}{\textbf{Function-level}} \\
  \cmidrule{3-14}
  && \textbf{Recall@1}   & \textbf{Recall@3}   & \textbf{Recall@5} & \textbf{MAP} & \textbf{MRR} & \textbf{nDCG@5}  & \textbf{Recall@1}   & \textbf{Recall@3}   & \textbf{Recall@5}  & \textbf{MAP} & \textbf{MRR} & \textbf{nDCG@5} \\
  \midrule
  \multicolumn{14}{c}{\cellcolor{gray!11} \textbf{Proprietary Models}} \\
  \midrule
  Agentless & GPT-4o & 62.28 & 79.41 & 85.01 & 73.02 & 75.72 & 76.83 & 28.50 & 43.56 & 46.55 & 37.17 & 43.32 & 41.45 \\
  Agentless & Claude-3.7-Sonnet & 66.84 & 82.48 & 86.76 & 76.47 & 79.76 & 79.97 & 36.09 & 51.21 & 53.83 &  45.12 &  52.54 & 49.49 \\
  CoSIL & Claude-3.7-Sonnet & \textbf{\underline{70.69}} & 86.07 & 89.13 & \textbf{\underline{79.73}} & \textbf{\underline{82.87}} & \textbf{\underline{82.96}} & \textbf{\underline{50.01}} & \textbf{61.56} & \textbf{66.38} & \textbf{58.43} & \textbf{67.28} & \textbf{63.17} \\
  LocAgent & Claude-3.7-Sonnet & 68.58 & 84.36 & 88.53 & 77.64 & 79.57 & 79.95 & 41.75 & 53.67 & 60.59 & 47.25 & 51.36 & 51.85 \\ 
  {RepoSearcher} & {Claude-3.7-Sonnet} & {69.48} & \textbf{\underline{87.09}} & \textbf{\underline{89.24}} & {79.39} & {82.31} & {82.73} & {48.55} & {61.57} & {66.08} & {57.76} & {65.94} & {62.53} \\
  \midrule
  \multicolumn{14}{c}{\cellcolor{gray!11} \textbf{Open-source 7B Models}} \\
  \midrule
  Agentless & Qwen-7B & 47.81 & 61.59 & 63.76 & 55.09 & 57.85 & 58.01 & 19.73 & 24.12 & 25.10 & 22.29 & 27.16 & 24.23 \\
  OrcaLoca & Qwen-7B & 47.69 & 50.29 & 50.29 & 49.09 & 52.40 & 50.14 & 16.44 & 27.23 & 28.57 & 22.51 & 26.41 & 25.27 \\
  LocAgent & Qwen-7B & 46.78 & 58.98 & 60.40 & 53.11 & 56.20 & 55.76 & 10.91 & 16.72 & 17.72 & 13.93 & 16.50 & 15.59 \\
  LocAgent & CL-7B & 56.11 & 72.11 & 75.18 & 65.68 & 68.61 & 68.87 & 24.82 & 30.29 & 34.12 & 27.81 & 33.56 & 32.99 \\
  \midrule
  CoSIL & Qwen-7B & 51.11 & 65.28 & 69.93 & 59.43 & 62.74 & 62.95 & 23.37 & 29.48 & 30.47 & 26.95 & 32.44 & 29.27 \\
  \rowcolor{blue!8} CoSIL & ToolTrain-7B & 53.13 & 67.35 & 72.09 & 61.54 & 64.33 & 64.95 & 34.86 & 46.01 & 50.87 & 42.49 & 48.44 & 46.50  \\
  \midrule  
  {RepoSearcher} & {Qwen-7B} & {38.33} & {53.23} & {57.73} & {46.36} & {49.20} & {49.95} & {18.12} & {24.54} & {25.82} & {21.60} & {24.91} & {23.53} \\
  \cellcolor{blue!8}{RepoSearcher} & \cellcolor{blue!8}{ToolTrain-7B} & \cellcolor{blue!8}{\textbf{59.78}} & \cellcolor{blue!8}{\textbf{77.89}} & \cellcolor{blue!8}{\textbf{83.11}} & \cellcolor{blue!8}{\textbf{70.43}} & \cellcolor{blue!8}{\textbf{73.46}} & \cellcolor{blue!8}{\textbf{74.44}} & \cellcolor{blue!8}{\textbf{43.59}} & \cellcolor{blue!8}{\textbf{57.32}} & \cellcolor{blue!8}{\textbf{62.38}} & \cellcolor{blue!8}{\textbf{53.28}} & \cellcolor{blue!8}{\textbf{60.44}} & \cellcolor{blue!8}{\textbf{57.91}} \\
  \midrule
  \multicolumn{14}{c}{\cellcolor{gray!11} \textbf{Open-source 32B Models}} \\
  \midrule
  Agentless & Qwen-32B & 58.56 & 74.67 & 79.50 & 68.03 & 71.14 & 71.75 & 26.30 & 38.85 & 40.70 & 33.32 & 39.44 & 37.03 \\
  OrcaLoca & Qwen-32B & 56.78 & 61.21 & 61.21 & 59.16 & 63.43 & 60.63 & 19.94 & 38.91 & 40.94 & 30.20 & 35.57 & 34.69 \\
  LocAgent & Qwen-32B & 64.15 & 77.12 & 78.02 & 71.44 & 74.13 & 73.79 & 20.97 & 40.00 & 45.34 & 31.61 & 36.47 & 36.61 \\
  LocAgent & CL-32B & 67.99 & 84.29 & 87.66 & 73.04 & 76.09 & 75.31 & 39.41 & 52.49 & 59.69 & 46.07 & 50.22 & 50.80 \\
  \midrule
  CoSIL & Qwen-32B & 59.34 & 78.73 & 82.68 & 69.88 & 73.02 & 73.92 & 38.62 & 50.27 & 54.41 & 45.96 & 53.94 & 50.34 \\
  \rowcolor{blue!8} CoSIL & ToolTrain-32B & 65.61 & 83.81 & 88.07 & 76.87 & 79.54 & 80.68 & 47.42 & 60.85 & 64.22 & 56.29 & 64.53 & 60.80 \\
  \midrule
  RepoSearcher & Qwen-32B & 58.79 & {76.33} & {81.03} & {68.77} & {71.72} & {72.67} & {32.38} & {44.35} & {48.19} & {39.45} & {45.89} & {43.50} \\
  \cellcolor{blue!8}{RepoSearcher} & \cellcolor{blue!8}{ToolTrain-32B} & \cellcolor{blue!8}\textbf{68.03} & \cellcolor{blue!8}\textbf{85.31} & \cellcolor{blue!8}\textbf{88.59} & \cellcolor{blue!8}\textbf{78.20} & \cellcolor{blue!8}\textbf{80.86} & \cellcolor{blue!8}\textbf{81.60} & \cellcolor{blue!8}\textbf{49.94} & \underline{\cellcolor{blue!8}\textbf{64.26}} & \underline{\cellcolor{blue!8}\textbf{68.55}} & \underline{\cellcolor{blue!8}\textbf{59.87}} & \cellcolor{blue!8}\textbf{\underline{67.35}} & \underline{\cellcolor{blue!8}\textbf{64.64}} \\
  \bottomrule
  \end{tabular}
  }
  \label{tab:rq1}
\end{table*}

\subsection{Effectiveness}
\paragraph{\textbf{Setup}}
To evaluate the performance of \textit{RepoSearcher} with the ToolTrain-model on the issue localization task, we compared it with four different issue localization frameworks (Agentless, Orcaloca, CoSIL, and LocAgent) using different models (Qwen-7B, Qwen-32B, and Claude-3.7-Sonnet). We conducted a multi-dimensional assessment of the various methods in terms of file localization and function localization, including multiple evaluation metrics: recall@1, recall@3, recall@5, MAP, MRR, and nDCG@5.
\paragraph{\textbf{Results}}
As shown in Table~\ref{tab:rq1}, \textit{RepoSearcher} with ToolTrain-model achieves state-of-the-art (SOTA) performance among same-size models and even surpasses leading commercial models on some function-level metrics. Although the \textit{RepoSearcher} underperformed other agent frameworks when using the original Qwen models due to its simpler agent design, after applying \textit{ToolTrain}, it achieves SOTA performance among same-size models. For example, \textit{RepoSearcher} with ToolTrain-32B achieves 68.55 at function-level Recall@5. Furthermore, \textit{RepoSearcher} with ToolTrain-model also demonstrates superior performance compared to CL-7B and CL-32B, which were trained specifically for the LocAgent. This demonstrates that \textit{ToolTrain} can effectively enhance the issue localization capability of LLMs and offers advantages over previous training approaches.

On function-level localization, \textit{RepoSearcher} with ToolTrain-7B even outperforms other frameworks that use 32B models. This indicates that training with ToolTrain effectively enhances an LLM's tool-calling capabilities, allowing a model with a smaller parameter count to achieve better performance than a much larger one.
It is also noteworthy that \textit{RepoSearcher} with ToolTrain-32B achieves performance on issue localization comparable to Claude-3.7-Sonnet. It even surpasses the commercial model on several fine-grained, function-level metrics (e.g., 68.55 v.s. 66.38 on function-level Recall@5). This detailed level of accuracy is potentially more impactful for downstream issue resolution tasks. This suggests that \textit{RepoSearcher} with ToolTrain-model has the potential to serve as a lightweight alternative to proprietary models like Claude~\citep{claude37}.

To validate the generalization ability of the ToolTrain-model, we also evaluated its performance on the CoSIL benchmark. As shown in Table~\ref{tab:rq1}, the ToolTrain-model achieves better performance on CoSIL compared to the original Qwen-model (e.g., 50.87 v.s. 30.47 on function-level Recall@5). This indicates that the ToolTrain-model, trained on \textit{RepoSearcher}, can successfully generalize its issue localization capabilities to other agents.

\begin{figure*}[t]
  \centering
  \begin{subfigure}[b]{0.99\textwidth}
    \centering
    \includegraphics[width=\textwidth]{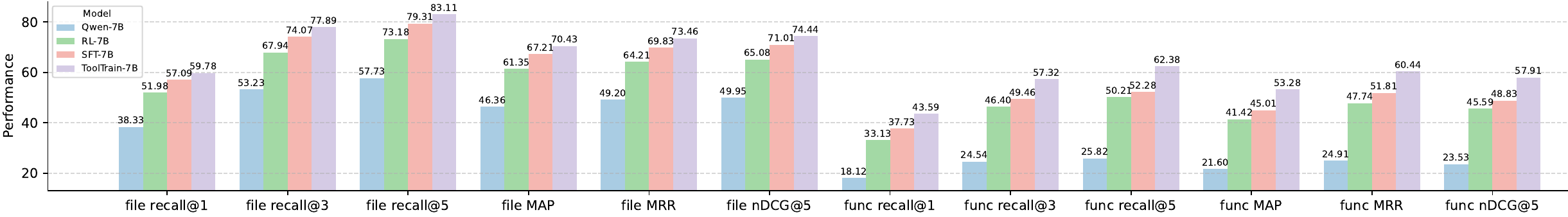}
    \caption{Training strategy ablation on 7B models.}
    \label{fig:ablation-7b}
  \end{subfigure}
  \hfill
  \begin{subfigure}[b]{0.99\textwidth}
    \centering
    \includegraphics[width=\textwidth]{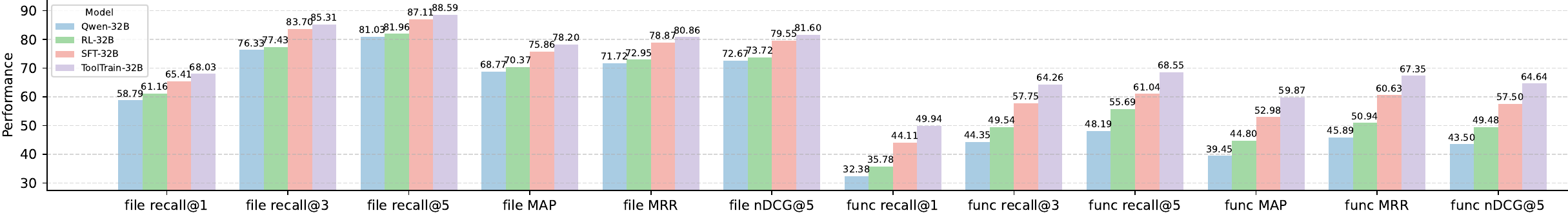}
    \caption{Training strategy ablation on 32B models.}
    \label{fig:ablation-32b}
  \end{subfigure}
  \caption{Training strategy ablation.}
  \label{fig:ablation}
\end{figure*}

\subsection{Training Strategy Ablation}
\paragraph{\textbf{Setup}}
To evaluate the effectiveness of \textit{ToolTrain}, we compared it with two other training strategies: (1) SFT: Directly performing Supervised Fine-Tuning on the LLM using the trajectories sampled in the first stage of \textit{ToolTrain}. (2) RL: Directly applying tool-integrated reinforcement learning to the LLM without any prior SFT.
We applied these three training strategies to both 7B and 32B models. We then evaluated the issue localization performance of the different models on the \textit{RepoSearcher} framework.
\paragraph{\textbf{Results}} As shown in Figure~\ref{fig:ablation}, \textit{ToolTrain} model outperforms both SFT and RL models of the same parameter size in issue localization.
Although training an LLM with either RL or SFT alone effectively improves issue localization performance, \textit{ToolTrain}'s combination of the two further boosts the LLM's performance across all issue localization metrics. For example, the base 7B model's file-level recall@5 is 57.73. This increases to 73.18 for RL-7B and 79.31 for SFT-7B, while ToolTrain-7B pushes it even further to 83.11.

Figure~\ref{fig:ablation} also indicates that standalone SFT yields a greater improvement than standalone RL. This can be attributed to the learning dynamics of each method. The RL process requires the model to first discover how to properly invoke tools through extensive exploration, leaving less capacity to refine its reasoning policies. SFT, on the other hand, explicitly provides examples of both correct tool usage and effective reasoning strategies, allowing the model to learn more efficiently. Consequently, standalone SFT delivers a larger improvement in issue localization.

\begin{table}
  \centering
  \caption{Issue resolution performance of different issue localization frameworks.}
  \label{tab:RQ3}
  \resizebox{.7\linewidth}{!}{
      \begin{tabular}{c|c|c|cc}
      \hline
      \textbf{Framework} & \textbf{Loc Model} & \textbf{Patch Model} & \textbf{Func Recall@5} & \textbf{\%Resolved} \\
      \hline
      Agentless & Qwen-7B & Qwen-7B & 25.10 & 7.60 \\
      CoSIL & Qwen-7B & Qwen-7B & 30.47 & 10.80 \\
      \rowcolor{blue!8} RepoSearcher & ToolTrain-7B & Qwen-7B & \textbf{62.38} &	\textbf{14.00} \\
      \hline
      Agentless & Qwen-32B & Qwen-32B & 40.70 & 25.80 \\
      CoSIL & Qwen-32B & Qwen-32B & 54.41 & 26.40 \\     
      \rowcolor{blue!8} RepoSearcher & ToolTrain-32B & Qwen-32B & \textbf{68.55} &	\textbf{31.60} \\
      \hline
      
      \end{tabular}
  }
\end{table}


\subsection{Influence on issue resolution}
\paragraph{\textbf{Setup}}
To evaluate the impact of different issue localization results on the final issue resolution performance, we use the localization results from various methods on SWE-Bench-Verified as input. We then use the patch generation script of agentless to generate patches. For localization methods employing 7B models, we use Qwen-7B as the patch generation model, and for those using 32B models, we use Qwen-32B. Following the standard evaluation method of SWE-Bench-Verified, a patch is considered to have successfully resolved the corresponding issue if it passes all unit tests. Based on this, we assess the impact of different localization methods on issue resolution using the issue resolution rate (\%Resolved). 
\paragraph{\textbf{Results}} As shown in Table~\ref{tab:RQ3}, more precise issue localization leads to better issue resolution results. \textit{RepoSearcher} with ToolTrain-32B achieves a function-level localization recall@5 of 68.55, and when using Qwen-32B as the patch generation model, it attains an issue resolution rate of 31.60, the best among all methods. Similarly, \textit{RepoSearcher} with ToolTrain-7B achieves a function-level localization recall@5 of 62.38, and with Qwen-7B as the patch generation model, its issue resolution rate is 14.00, marking the best performance among the 7B models.
The table also reveals that despite the comparable localization results between \textit{RepoSearcher} with ToolTrain-32B and \textit{RepoSearcher} with ToolTrain-7B, the use of different patch generation models results in a significant disparity in their resolution outcomes (14.00 vs. 31.60). This indicates that the capability of the patch generation model is also crucial for issue resolution. Therefore, future work should focus on further enhancing the patch generation capabilities of models.

\section{Related Work}\label{sec:related}

\subsection{Fault Localization}
Fault localization refers to locating a faulty code snippet within a repository based on a failed test~\citep{raselimo2019spectrum, abreu2009spectrum, qin2025s, wen2019historical} or an issue description~\citep{jimenez2024swebench}. DeepFL~\citep{li2019deepfl} utilizes multiple fault-diagnosis dimensions, and combines them with a deep neural network to localize faulty code. DeepRL4FL~\citep{li2021fault} works by encoding the test coverage matrix into a feature matrix. It then integrates this with inter-statement data dependencies and static representations of the code, using a Convolutional Neural Network (CNN)~\citep{li2021survey} to automatically identify and locate buggy code statements or methods.

With the advancement of large language models in language understanding and reasoning, an increasing number of studies have started focusing on fault localization based on issue description. These range from lightweight pipelines to autonomous agents. Agentless \citep{agentless} represents the pipeline approach, using a hierarchical strategy that combines LLMs and retrievers to progressively narrow down from files to fine-grained code locations. In parallel, agent-based systems have gained traction. A key strategy is representing code as a graph to guide the agent. LocAgent \citep{chen2025locagent} preprocesses the repository into a comprehensive static graph to enable efficient multi-hop reasoning. In contrast, CoSIL \citep{jiang2025cosilsoftwareissuelocalization} dynamically constructs module call graphs during its search, enabling iterative exploration. Another line of research, exemplified by OrcaLoca \cite {yu2025orcalocallmagentframework}, focuses on enhancing the agent's internal mechanics through priority-based action scheduling, action decomposition, and distance-aware context pruning to boost accuracy. In this paper, we introduce \textit{RepoSearcher}, a lightweight issue localization agent. Its lightweight design enables the rapid sampling of numerous trajectories, allowing us to efficiently train an LLM specifically tailored for the agent.

\subsection{Agentic Training}

Although Large Language Models (LLMs) have achieved near-human performance in many general domains, their performance in issue resolution as agents still lags significantly behind that of humans due to a lack of agent-specific training. Many works have attempted to post-train LLMs for agent-oriented tasks to enhance their ability to use tools, reason, and complete tasks. The most critical challenge in this process is ensuring the quality of the training data.
To address this, SWE-Gym~\citep{pantraining} constructs an environment of real Python repositories, executable environments, and unit tests. It filters for high-quality agent trajectories based on the outcomes of unit test executions and uses these trajectories to supervised fine-tune the LLM, thereby enhancing its agent capabilities. Similarly, SEAlign~\citep{zhang2025sealign} proposes an alignment training framework for software engineering agents. It gathers high-quality engineering process trajectories, employs Monte Carlo Tree Search (MCTS)~\citep{swiechowski2023monte} for fine-grained scoring during multi-step decision-making, and combines this with preference optimization on key actions to supervise the fine-tuning of the LLM.
LocAgent~\citep{chen2025locagent} also constructs the ground truth for issue localization based on functions modified by \textit{golden patches} from GitHub. This is used to filter out low-quality trajectories sampled by LocAgent to ensure the quality of the training data. While the training data in prior work was mostly built upon existing issues, SWE-smith~\citep{yang2025swe} first establishes an execution environment from any Python repository and then generates a large number of task instances that break existing tests. This allows for the efficient synthesis of a dataset with hundreds of thousands of tasks, each with execution-based validation.
While previous work has primarily used SFT to train LLMs~\citep{chen2025locagent,yang2025swe}, in this paper, we explore how to organically combine SFT and RL to further enhance the LLM's issue localization capabilities. To the best of our knowledge, we are the first work to organically combine SFT and RL for training an issue localization agent.

\section{Conclusion}\label{sec:conclusion}

In this paper, we improve the issue localization performance of LLMs by introducing \textit{ToolTrain}, a tool-integrated training approach to enhance their tool-use and reasoning. 
Our framework first utilizes supervised fine-tuning to warm up the model with our lightweight agent, \textit{RepoSearcher}, and then employs tool-integrated reinforcement learning to teach the model how to effectively navigate code repositories using various tools.
We apply \textit{ToolTrain} to open-source models, and evaluate them on issue localization tasks. The experiment results show that \textit{ToolTrain} trained model achieve state-of-the-art performance among same-size models, and even surpassing leading models like Claude-3.7 on specific tasks. The code is available at \url{https://github.com/Mizersy/RepoDeepSearch}

\section*{Acknowledgments}
This work is supported by National Key Research and Development Program of China (Grant No. 2023YFB4503803).

\bibliography{reference}
\bibliographystyle{reference}

\newpage
\appendix
\section*{Appendix}

\section{Tool Call Analysis}

\begin{figure}[h]
  \centering
  \includegraphics[width=.6\textwidth]{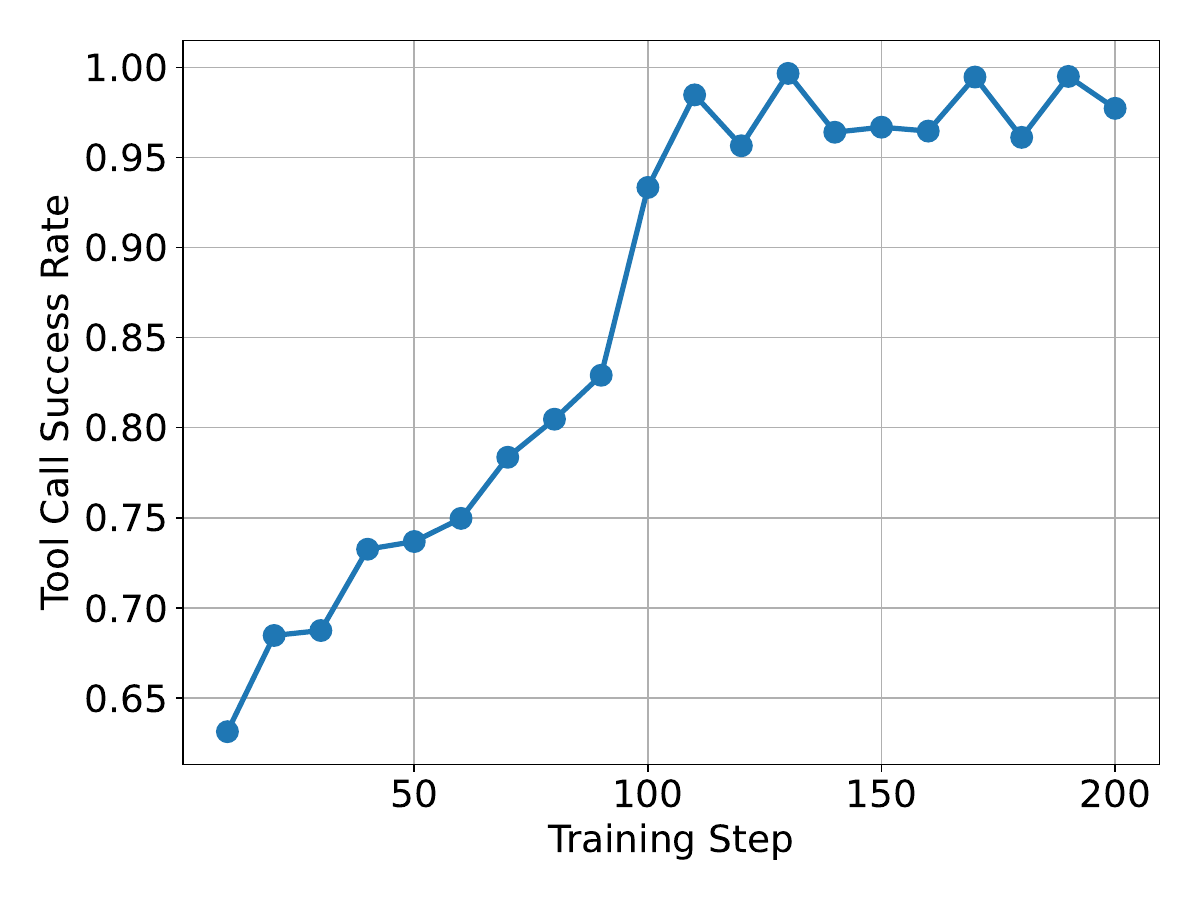}
  \vspace{-3mm}
  \caption{Tool call success rate during RL training.}
  \label{fig:toolAccCurve}
  \vspace{-3mm}
\end{figure}

\textbf{Tool-integrated reinforcement learning can effectively improve the tool call success rate of LLMs.} We analyzed the progression of the LLM's tool call success rate during the reinforcement learning (RL) training process on 7B model. A tool call is deemed successful when it invokes an existing tool with valid parameters (such as filenames and function names that are present in the repository) and successfully retrieves a result from it. As shown in Figure~\ref{fig:toolAccCurve}, the LLM's tool call success rate exhibits a steady increase throughout the training, eventually stabilizing at over 95\%. This indicates that in tool-integrated reinforcement learning, even when the reward is calculated solely based on the correctness of the final localization result, the model's proficiency in tool calling can be effectively enhanced. A plausible explanation is that during the RL process, the LLM learns to call tools with greater efficiency and accuracy in order to more effectively acquire the necessary issue-related project context.

\section{Case study}
\begin{figure*}[h]
  \centering
  \includegraphics[width=.99\textwidth]{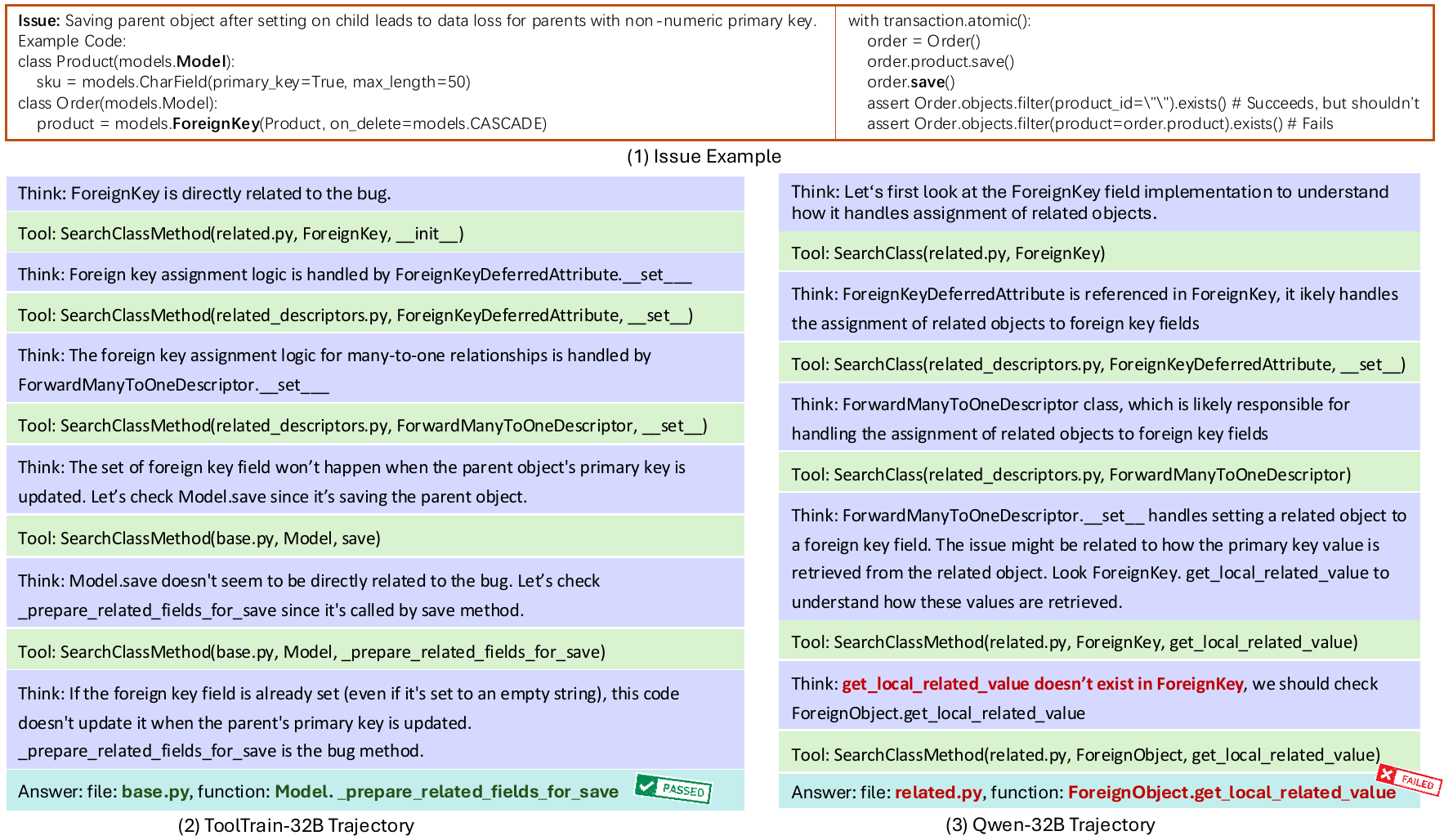}
  \caption{Case study.\protect\footnotemark}
  \label{fig:case}
\end{figure*}

The case in Figure~\ref{fig:case} demonstrates that \textit{ToolTrain} can effectively enhance the tool-calling and reasoning abilities of LLMs in the issue localization process.

For the issue presented in Figure~\ref{fig:case}(1), we conducted a detailed analysis of the reasoning and tool-calling performance of ToolTrain-32B and Qwen-32B during the issue localization process.
As shown in Figure~\ref{fig:case}(2), ToolTrain-32B demonstrates the ability to reason and call tools accurately during localization. It first identifies that the ``ForeignKey'' field is directly related to the bug and progressively traces the issue to key methods such as ForeignKeyDeferredAttribute.\_\_set\_\_ and ForwardManyToOneDescriptor.\_\_set\_\_. Upon realizing that the functions along this path are not the root cause of the bug, it precisely pivots its search direction, ultimately locating the Model.\_prepare\_related\_fields\_for\_save function in base.py. Throughout this process, ToolTrain-32B not only shows a strong ability to comprehend the problem but also exhibits excellent inter-module dependency analysis skills. It proactively inspects methods related to primary key updates and foreign key assignments, reasonably utilizes search tools to uncover the evidence path, and ultimately succeeds in locating the bug's source.
In contrast, Qwen-32B, as shown in Figure~\ref{fig:case}(3), exhibits significant shortcomings. Although it also identified ForeignKey-related classes and methods in the initial stage, its search scope remained confined to a few explicitly mentioned fields and classes. It failed to break out of this initial path to perform broader reasoning about the code structure, ultimately leading it to incorrectly pinpoint ForeignObject.get\_local\_related\_value and causing the task to fail. Furthermore, Qwen-32B made an incorrect tool call by attempting to retrieve a non-existent method ForeignKey.get\_local\_related\_value.
Ultimately, this example validates the effectiveness of our \textit{ToolTrain} framework in cultivating the deep and flexible reasoning, which is necessary to navigate complex codebases and overcome challenging localization tasks.
\footnotetext{Due to space limitations, the issue content and think content have been simplified. The original issue content can be found at: https://code.djangoproject.com/ticket/32332}

\section{Threats to Validity}\label{sec:threats}
A limitation of our evaluation lies in the construction of the ground truth. For the issue localization task, we define the ground truth as the files and functions modified by the provided ``golden patch.'' However, the solution to a given issue is not always unique, meaning other code locations could potentially serve as an equally valid fix. Consequently, our evaluation methodology may not credit these alternative, correct solutions.
Nevertheless, we maintain that using the golden patch as a benchmark is an effective means of evaluating and comparing different methods. The primary goal of our task is to identify suspicious files and functions. A model that successfully pinpoints a location known to be correct (i.e., the one in the golden patch) is demonstrably more effective than one that does not. Therefore, this approach provides a consistent and practical standard for assessing the performance of issue localization frameworks.

The evaluation was conducted on Python repositories, and the performance on other languages remains to be validated. This is mainly because the evaluation set was constructed based on SWE-Bench-Verified, which is focused on the Python language.
In our future work, we plan to apply \textit{RepoSearcher} and \textit{ToolTrain} to more languages. This will allow us to further verify the effectiveness of our method and contribute to the advancement of automated issue localization.

\section{Details of Baseline Frameworks}\label{sec:detailsOfBaseline}
\paragraph{\textbf{Agentless}} Agentless~\citep{agentless} is a pipeline-based issue resolution framework. In the localization stage, it adopts a hierarchical localization strategy: it first identifies suspicious files in the repository using LLMs, then narrows down to relevant classes or functions, and finally pinpoints fine-grained edit locations by LLMs for precise issue localization.
\paragraph{\textbf{OrcaLoca}} Orcaloca~\citep{yu2025orcalocallmagentframework} is an agent-based issue localization framework that integrates priority-based LLM-guided action scheduling, action decomposition with relevance scoring, and distance-aware context pruning to enhance the agent's issue localization accuracy.
\paragraph{\textbf{CoSIL}} CoSIL~\citep{jiang2025cosilsoftwareissuelocalization} is a function-level software issue localization Agent that utilizes LLMs to dynamically construct module call graphs during the repository search process, iteratively explores relevant contexts, and employs context pruning to effectively narrow the search space.
\paragraph{\textbf{LocAgent}} LocAgent~\citep{chen2025locagent} is an issue localization agent that transforms repositories into directed heterogeneous graphs, capturing structures like files, classes, and functions, along with their dependencies such as imports and invocations. This representation allows LLM agents to perform efficient multi-hop reasoning, precisely identifying relevant code from natural language descriptions.

\section{Details of Metrics}\label{sec:detailsOfMetrics}

\paragraph{\textbf{Recall@k}} Recall@k measures the proportion of ground-truth answers that are included in the top-k localization results. Previous study have shown that 73.58\% of developers only examine the top-5 localization results~\citep{kochhar2016practitioners}. Therefore, following previous work~\citep{jiang2025cosilsoftwareissuelocalization}, we set the maximum value of k to 5 and evaluate the performance at Recall@1, Recall@3, and Recall@5.

\paragraph{\textbf{MAP}} MAP (Mean Average Precision) evaluates the overall ranking quality by averaging the \textbf{Average Precision (AP)} across all issues. For each issue, AP measures the average of the precision values at the ranks where relevant (ground-truth) items occur. This metric reflects both the presence and the ranking of relevant results, and is calculated as Equation~\ref{eq:ap}. Then, MAP is computed as Equation~\ref{eq:map}.
\begin{equation}
\text{MAP} = \frac{1}{|Q|} \sum_{q \in Q} \text{AP}(q),
\label{eq:map}
\end{equation}
\begin{equation}
\text{AP}(q) = \frac{1}{|A_q|} \sum_{k=1}^{|L_q|} \text{Precision@}k \cdot \mathbb{I}[L_q[k] \in A_q],
\label{eq:ap}
\end{equation}
where \( Q \) is the set of all queries. For a query \( q \in Q \), let \( A_q \) be the set of ground-truth relevant items.  $\text{AP}(q)$ refers to the \textbf{Average Precision} for query \( q \), \( L_q \) refers to the predicted ranked list for query \( q \), \( \text{Precision@}k \) is the precision at rank \( k \), \( \mathbb{I}[L_q[k] \in A_q] \) is an indicator function that is 1 if the item at rank \( k \) is relevant, and 0 otherwise.

\paragraph{\textbf{MRR}} MRR (Mean Reciprocal Rank) evaluates how early the ground-truth item appears in the ranked list. It is the average of the reciprocal ranks of the ground-truth results across all queries. The higher the MRR, the better the system is at placing the correct result near the top.
\begin{equation}
\text{MRR} = \frac{1}{|Q|} \sum_{q \in Q} \frac{1}{\text{rank}_q},
\end{equation}
where  \( \text{rank}_q \) is the rank position of the ground-truth result for query \( q \).

\paragraph{\textbf{nDCG@$k$}} The $\text{nDCG@}k$  (Normalized Discounted Cumulative
Gain at rank k) score is computed as the mean $\text{nDCG@}k(q)$ over all queries:
\begin{equation}
\text{nDCG@}k = \frac{1}{|Q|} \sum_{q \in Q} \text{nDCG@}k(q).  
\end{equation}

\paragraph{\textbf{\%Resolved}} The \%Resolved is used to evaluate the accuracy of generated patch in RQ3. \%resolved calculates the proportion of generated patches that pass all unit tests.
\end{document}